\documentclass[12pt]{article}
\pdfoutput=1
\usepackage{geometry,enumerate,amsmath,amssymb}
\usepackage{fullpage}
\usepackage{graphicx}
\usepackage{bm}
\numberwithin{equation}{section}

\newcommand{\be}{\begin{equation}}
\newcommand{\ee}{\end{equation}}
\newcommand{\bea}{\begin{eqnarray}}
\newcommand{\eea}{\end{eqnarray}}

\renewcommand{\epsilon}{\varepsilon}

\newcommand{\bnabla}{\mbox{\boldmath $\nabla$}}

\begin{document}
\title{
  Rings on strings in excitable media
}
\author{
  Fabian Maucher$^{\dagger\star}$ and Paul Sutcliffe$^\star$\\[10pt]
{\em \normalsize
  $^\dagger$Joint Quantum Centre (JQC) Durham-Newcastle, Department of Physics,}
  \\
{\em \normalsize  Durham University, Durham DH1 3LE, United Kingdom.}\\
 {\em \normalsize $^\star$Department of Mathematical Sciences,}\\
 {\em \normalsize Durham University, Durham DH1 3LE, United Kingdom.}\\ 
{\normalsize Email: fabian.maucher@durham.ac.uk, \ p.m.sutcliffe@durham.ac.uk}
}
\date{October 2017}

\maketitle
\begin{abstract}
  We study the dynamics and interaction of coaxial vortex rings in the FitzHugh-Nagumo excitable medium. We find that threading vortex rings with a vortex string results in significant qualitative differences in their evolution and interaction. In particular, threading prevents the annihilation of rings in a head-on collision, allows generic ring overtaking, and can even reverse the direction of motion of a ring. We identify that an important mechanism for producing this new behaviour is that threaded vortex rings interact indirectly via induced twisting of the threading vortex string. 
   \end{abstract}

\newpage
\section{Introduction}\quad
Spiral wave vortices have been extensively studied experimentally, modelled mathematically and simulated numerically in a range of biological, chemical and physical excitable media. Notable examples include
the chemotaxis of slime mould, oxidation waves in redox reactions, and depolarization waves in cardiac tissue \cite{Winfree,Win2}. In a three-dimensional excitable medium vortices form extended vortex strings that either end on the boundary of the medium or form closed loops. In the latter case a vortex ring is the simplest possibility.

The numerical study of vortex rings, and more general knotted and linked vortex strings, began more than thirty years ago in a series of papers by Winfree and Strogatz \cite{WinStr1,WinStr2,WinStr3,WinStr4}. Advances in computing power allowed the study of linked and knotted vortex strings over much longer time scales than early investigations and provided further evidence for their stability \cite{SutWin}. Modern parallel computations, together with a new technique for the initialization of vortex strings with arbitrary conformation and topology, has enabled a very recent detailed study of a range of different knot types \cite{MS2} and has revealed remarkably complex dynamics that includes untangling knots without untying \cite{MS1}. Naively, knot untangling might simply be understood as a consequence of vortex core repulsion, together with shrinking string length as captured by effective models built on local curvature driven dynamics \cite{Ke,BHZ}. However, it turns out that this simple picture is too crude due to the significant role that twist plays in the dynamics of vortex strings and the fact that linking enforces twisting \cite{WWS}.

The ultimate aim is to fully understand the mechanisms that produce complicated knot evolution and untangling. Here we make some progress towards this goal by studying simple geometries and identifying dynamical features of vortex strings that are topological, in the sense that they appear because of linking and the associated mandatory twist that this entails.
In detail, we investigate the dynamics of multiple coaxial vortex rings and contrast the evolution with the same situation modified by the addition of a single vortex string that threads all the rings. By comparing the findings both with and without the threading vortex string, we are able to report on the additional features that appear in the dynamics and interaction of twisted vortex rings due to their linking with the vortex string.

We perform our numerical simulations within the FitzHugh-Nagumo medium \cite{FH,Nag}, which is the simplest model used to describe cardiac tissue as an excitable medium \cite{Kog}. In section \ref{sec:one} we review the FitzHugh-Nagumo model and the dynamics of an isolated vortex ring solution. In section \ref{sec:unthreaded} we study the dynamics of a pair of coaxial vortex rings and observe that the generic outcome is the annihilation of at least one of the rings, usually via a swallowing phenomenon in which a smaller ring attempts to pass through a larger ring but is destroyed in the process. Finally, in section \ref{sec:threaded} we introduce an additional vortex string that threads all the coaxial vortex rings and find that this yields significant qualitative differences in the evolution and interaction of the vortex rings. In particular, annihilation is no longer the generic outcome and instead rings can overtake each other without one being swallowed by the other. This leads to a much richer dynamics for coaxial rings and includes new phenomena, such as the reversal of the direction of motion of a ring. We identify that an important mechanism for producing this new behaviour is that the vortex rings interact indirectly via induced twisting of the threaded vortex string.

\section{The vortex ring in the FitzHugh-Nagumo model}\quad\label{sec:one}
The FitzHugh-Nagumo medium is described by the nonlinear
reaction-diffusion equations 
\be
\frac{\partial u}{\partial t}=\frac{1}{\epsilon}(u-\frac{1}{3}u^3-v)+\nabla^2 u,
\quad\quad
\frac{\partial v}{\partial t}=\epsilon(u+\beta-\gamma v),
\label{FHN}
\ee
where $u({\bf r},t)$, represents the electric potential and
$v({\bf r},t)$ is the recovery variable, both being real-valued physical fields defined throughout the three-dimensional medium with spatial coordinate ${\bf r}$ and time $t$.
The Greek letters in the FitzHugh-Nagumo equation (\ref{FHN}) are constant parameters, which we take to be 
$\epsilon=0.3, \ \beta=0.7, \ \gamma=0.5$. This choice avoids complications
due to spiral wave meander \cite{Win2}.
With this choice of parameter values, the FitzHugh-Nagumo equation
has a two-dimensional rotating spiral wave vortex solution \cite{Winfree},
with a period ${T}=11.2$.  The vortex has $u$ and $v$ wavefronts in the form of an involute spiral with a wavelength $\lambda=21.3,$ and the speed of plane waves is given by $c=\lambda/{T}=1.9.$ The constants $T$ and $\lambda$ are the
characteristic time and length scales of the excitable system.

To visualize spiral wave vortex strings, it is useful to introduce the quantity 
\begin{equation}
 {\bm B} = \bnabla u \times \bnabla v,
\end{equation} 
because in two spatial dimensions the centre of a spiral wave vortex
is the point at which $|{\bm B}|$ is maximal,
and this quantity is localized in the vortex core \cite{Win3}.
In a three-dimensional medium the spiral wave tip forms a curve that is either closed or ends at the medium boundary. This vortex string
can be visualized as the centreline of the tube obtained by plotting the isosurface $|{\bm B}|=0.1$.

To compute numerical solutions of the FitzHugh-Nagumo partial differential equations (\ref{FHN}) we employ standard methods, with time evolution performed using a fourth-order Runge-Kutta method with timestep $\Delta t=0.1$ and spatial derivatives calculated using the discrete cosine transform with a lattice spacing $\Delta x=0.5$. No-flux (Neumann) boundary conditions are employed on all boundaries of the medium.
To generate initial conditions for vortex strings we use the fact that the one-parameter family of FitzHugh-Nagumo fields
\be u=2\cos\varphi-0.4, \qquad v=\sin\varphi-0.4,
\label{loop}
\ee
where $\varphi\in[0,2\pi)$, provides an adequate approximation to the excitation-recovery loop of the excitable medium. Initial fields $u({\bf r},0)$ and $v({\bf r},0)$ are given by specifying a phase profile $\varphi({\bf r})$
  that generates vortex strings at the phase singularities where $\varphi$ is undefined (if such a point coincides with a simulation lattice point then it is sufficient to set the phase to zero at this lattice point).
  
\begin{figure}[ht]\begin{center}              
 \includegraphics[width=0.8\columnwidth]{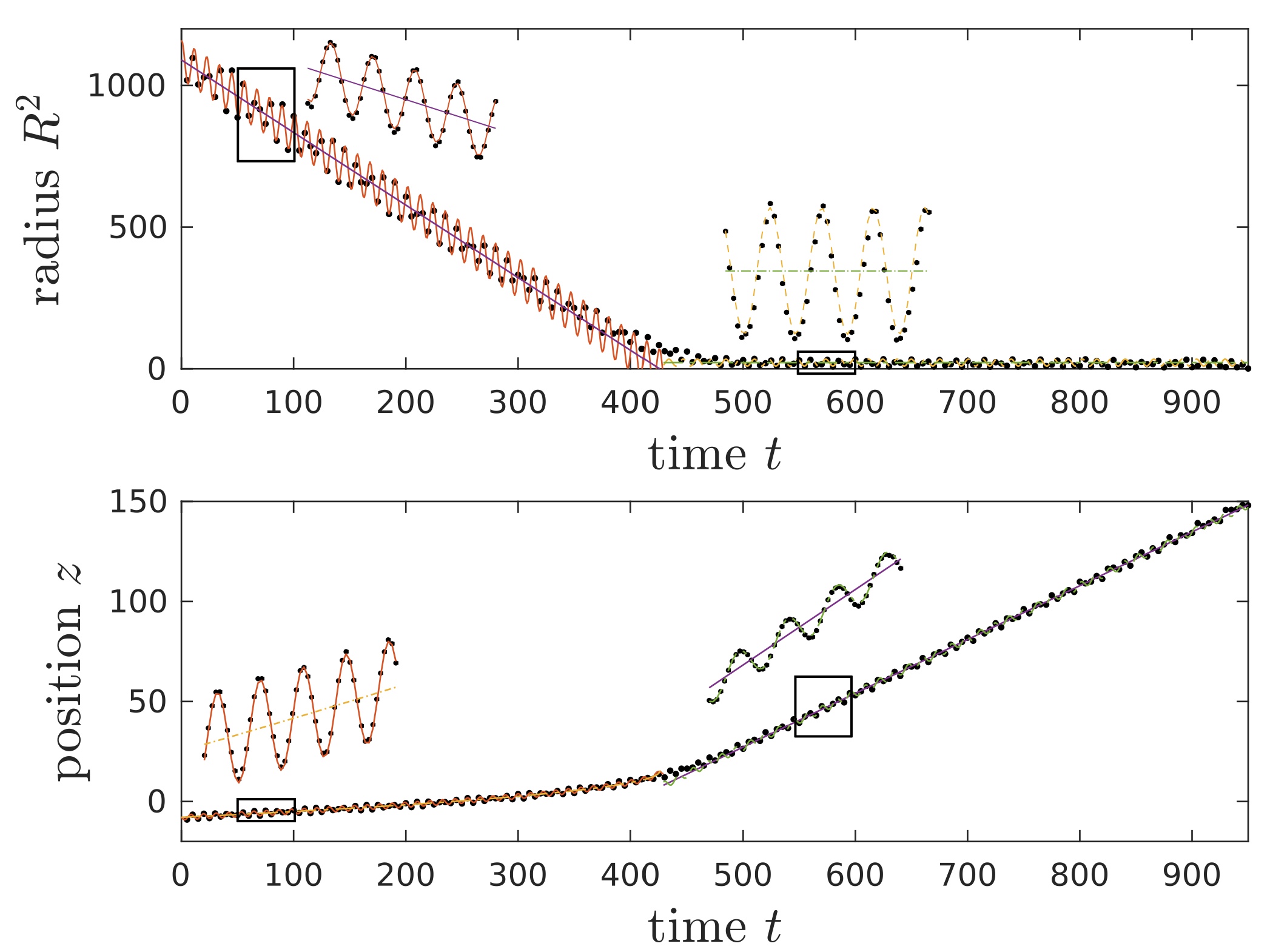}
 \caption{The results of a numerical simulation to study the dynamics of a single vortex ring with a large initial radius $R_0=33.$  
   The square of the radius is shown in the upper image and the position along the symmetry axis is shown in the lower image. The oscillating curves are fits to the numerical data, represented by dots, and the non-oscillating curves are the analytic approximations (\ref{onering}). The insets show magnified portions of the plots.
There are two regimes: one where the period is $T$ and the 
   radius shrinks, and one where the period is $1.14\,T$ and the radius oscillates about the value $R_\star=4.7$ with the ring drifting at a constant speed $0.27$.  }
\label{fig:onering}
\end{center}
\end{figure}

  To provide initial conditions for a vortex ring it is convenient to introduce cylindrical polar coordinates $\rho,\theta,z$, where $x=\rho\cos\theta$ and $y=\rho\sin\theta$. The phase profile for a vortex ring with initial radius $R_0$ and vertical position $z_0$ is given by the axially symmetric formula
 \be
 \varphi_{ring} = \arg\big(\rho-R_0\pm i(z-z_0)\big),
 \label{phi_ring}
\ee
where the choice of sign determines whether the ring moves up or down the $z$-axis. 
We begin by reviewing the well-known dynamics of an isolated vortex ring \cite{CSW}. The main features are presented in Fig.~\ref{fig:onering}, where we display the results of the evolution of the radius and position of a ring (obtained by locating the maximal value of $|{\bm B}|$) with a large initial radius $R_0=33$ and an initial position $z_0=-7.9.$
For the parameter values used in this paper there is a stable vortex ring with a radius $R_\star=4.7\approx \lambda/5$ that drifts along its symmetry axis like a smoke ring with a speed $0.27\approx c/7$.

For a vortex ring with an initial radius $R_0\gg R_\star$ the radius has small amplitude oscillations with the period $T$ but averaging over this period reveals the underlying evolution where the ring shrinks in a way that is well-described by curvature driven dynamics, with the rate of change of the radius being inversely proportional to the radius. This behaviour can be derived within an analytic approximation \cite{Ke,BHZ} that is valid for small curvature and within the regime that vortex cores are well-separated. A ring that shrinks in this way is usually used to define the system as one in which there is a positive filament tension. The motion along the symmetry axis can also be derived within the same approximation \cite{Ke,BHZ}. Orient the vortex ring so that the $z$-axis is the symmetry axis and define the initial position along this axis as $z_0$ and the subsequent position as $z(t).$ The small curvature approximation yields the results for the radius $R(t)$ and position $z(t)$ presented in the upper line in
(\ref{onering})
\be
 R(t)=
\begin{cases}
\sqrt{R_0^2-2c_1t} & \text{for } t<t_1\\
  R_\star  & \text{for } t\ge t_1
\end{cases}
, \quad
 z(t)=
\begin{cases}
z_0+\frac{c_2}{c_1}(R_0-\sqrt{R_0^2-2c_1t})  & \text{for } t<t_1 \\
  \widetilde z_0+\frac{c_3}{R_\star}t & \text{for } t\ge t_1
\end{cases}
\label{onering}
\ee
valid for $t<t_1=\frac{1}{2}(R_0^2-R_\star^2)/c_1$. Here the positive constant $c_1$ is, by definition, the filament tension and the positive constant $c_2$ is proportional to the drift speed of the ring along its symmetry axis. Both these constants depend upon the parameters in the FitzHugh-Nagumo equation (\ref{FHN}) and for the parameters used in this paper a fit yields the values $c_1=1.28$ and $c_2=0.89.$

For times greater than $t_1$ the diameter of the ring is smaller than the spiral wavelength and repulsive forces between the vortex cores prevents further shrinking, so that the radius is (almost) constant at the stabilized value $R_\star$. The radius is not perfectly constant because there is an oscillation associated with the relative position of the involute spiral wave and thus a small amplitude oscillation of the radius around $R_\star$, with a period
that is $14\%$ larger than the spiral period $T$.
In this regime, the expressions in the lower line in equations (\ref{onering}) are obtained from a constant radius approximation together with a linear fit for the time dependence of the position. With the parameter values $c_3=1.26$ and $\widetilde z_0=-107.2$ these expressions provide a good approximation to the late time evolution, with the ring drifting along its symmetry axis with the constant speed $c_3/R_\star=0.27$ mentioned earlier. Note that $c_3$ is larger than $c_2$, whereas simply fixing the radius by hand in the small curvature approximation would yield a drift speed with $c_3$ equal to $c_2.$

The numerical results and the analytic approximations are clearly within reasonable agreement for the motion of an isolated vortex ring, although even in this simple case we have seen that different approximations must be applied once vortex cores are no longer well-separated and local curvature driven motion fails to be an appropriate description.

\section{The interaction of coaxial vortex rings}\quad\label{sec:unthreaded}
The main objective of the present paper is to demonstrate that the presence of other vortex rings or strings significantly alters the dynamics of a vortex ring and that the topological property of linking or threading produces dramatic qualitative differences. In this section we discuss the evolution of rings in the absence of linking or threading and consider the simplest case of multiple rings, namely coaxial rings. In the subsequent section we shall then contrast these results with the same situation modified by the introduction of a single vortex string that threads all the coaxial vortex rings.

To produce initial conditions for coaxial vortex rings the phase profile is simply taken to be a sum of phase profiles of the single ring form (\ref{phi_ring}), with independent radii and positions for each ring. As mentioned earlier, we visualize vortex rings by plotting the isosurface $|{\bm B}|=0.1,$ which yields a torus for an isolated vortex ring. For later use, it is also helpful to colour this isosurface according to the value of the phase $\varphi$ defined by (\ref{loop}), namely,
\be
\varphi=\tan^{-1}\bigg(\frac{2(v+0.4)}{u+0.4}\bigg).
\ee
Any curve on the toroidal isosurface specified by a constant value of $\varphi$ has an integer linking number with the centre-line of the vortex ring. We refer to this linking number as the total twist and visually it is equal to the number of full turns made by the colour wheel as the toroidal direction of the vortex ring isosurface is traversed once. It is a consequence of a more general theorem \cite{WWS} that for each vortex ring the total twist (defined with an appropriate sign convention) is equal to the sum of the linking numbers of the given vortex ring with all the other rings or threading strings. As linking or threading is absent in this section, all rings will have zero total twist and indeed the colouring of any ring will remain fixed as one moves along the toroidal direction of the ring's isosurface, reflecting the fact that all the solutions discussed in this section are axially symmetric. Note that every colour on the colour wheel is attained once along the poloidal direction of the isosurface, because the phase singularity is located inside the torus. To help clearly identify and track individual rings during the evolution, we often include insets within each figure where each ring is coloured with a given fixed colour. 

\begin{figure}[ht]\begin{center}           
\includegraphics[width=0.81\columnwidth]{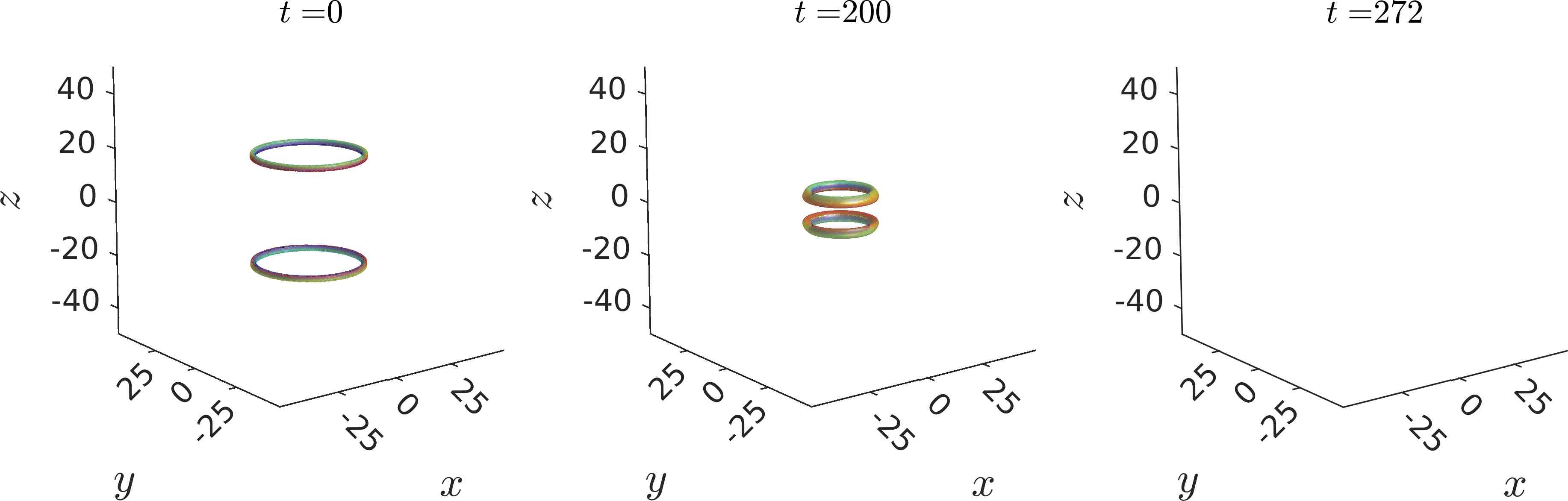}
\caption{Two coaxial rings moving in opposite directions collide head-on and mutually annihilate.
}
\label{fig:headon}
\end{center}\end{figure}
\begin{figure}[ht]\begin{center}           
\includegraphics[width=0.81\columnwidth]{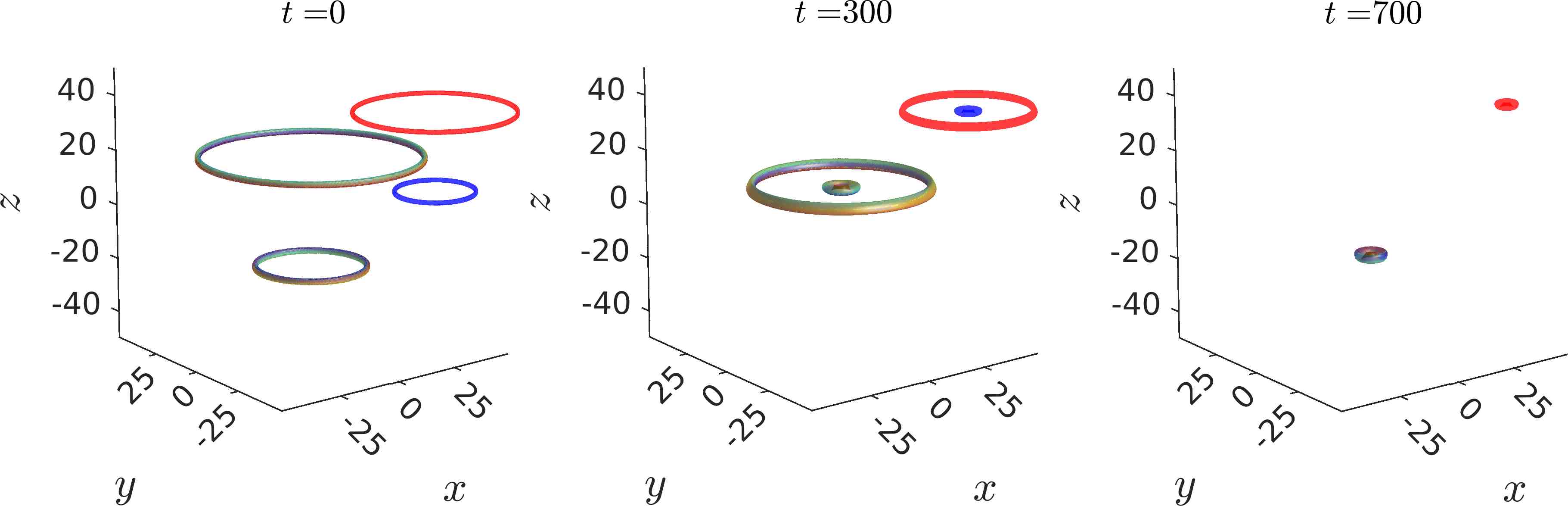}
\caption{Two coaxial rings with different initial radii move in opposite directions and collide head-on. The upper ring (red in inset) is larger that the lower ring (blue in inset). Both rings contract and the lower blue ring attempts to pass through the upper red ring. However, the blue ring is swallowed by the larger red ring which subsequently contracts to the stable ring radius.
  }
\label{fig:head_on_swallow}
\end{center}\end{figure}
\begin{figure}[ht]\begin{center}           
    \includegraphics[width=0.81\columnwidth]{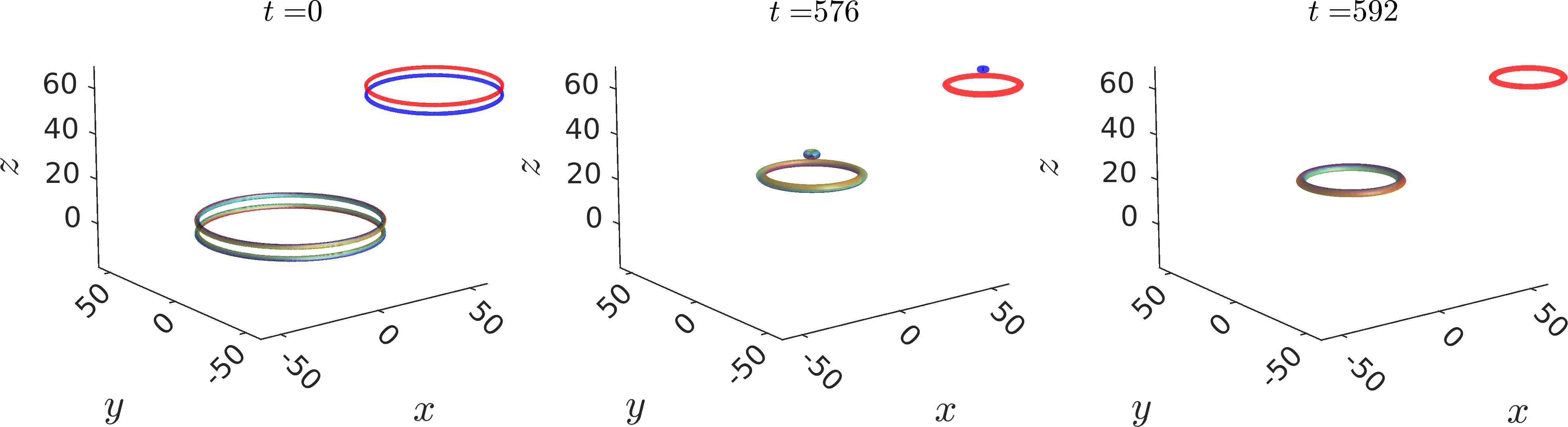}
\caption{Two coaxial rings moving in the same direction and initialized close together with large equal radii. The lower ring (blue in inset) shrinks and attempts to overtake the larger (and therefore slower) upper ring (red in inset). 
However, overtaking fails because the smaller blue ring is swallowed by the larger red ring before it can escape.
}
\label{fig:swallow}
\end{center}\end{figure}
\begin{figure}[ht]\begin{center}           
 \includegraphics[width=0.81\columnwidth]{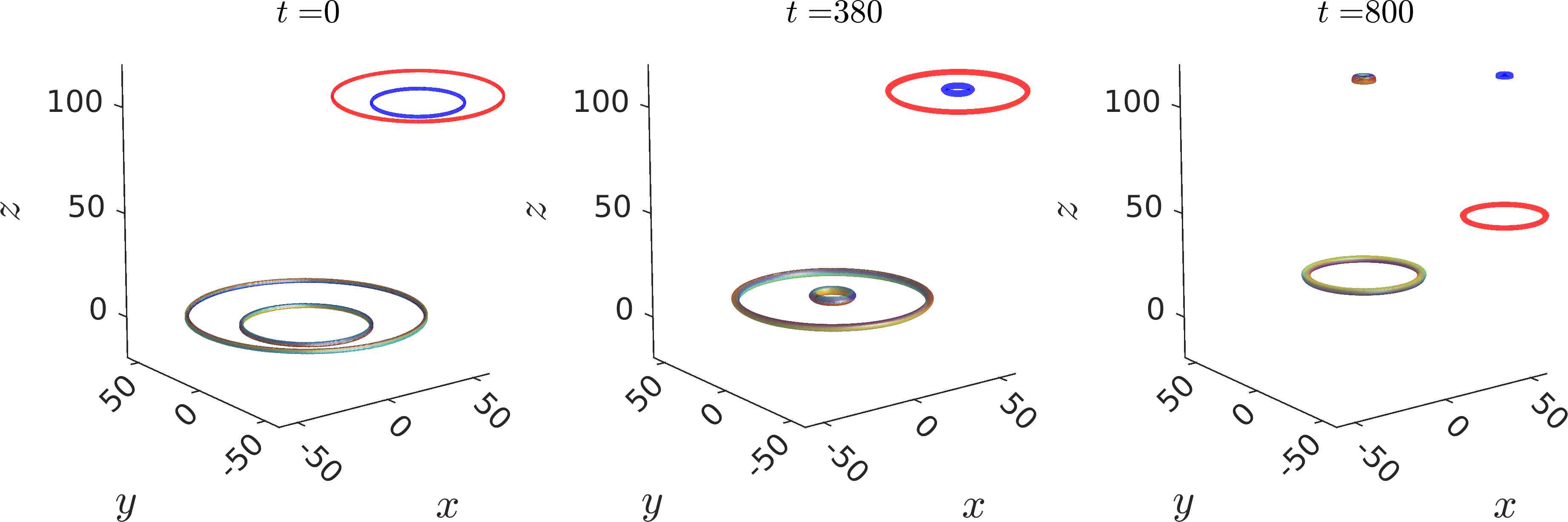}
 \caption{
Two coaxial rings moving in the same direction and initialized close together with the upper ring being larger than the lower ring. The lower ring (blue in inset) shrinks and overtakes the larger upper ring (red in inset) and escapes to produce two well-separated rings.   
}
\label{fig:overtake}
\end{center}\end{figure}
Fig.~\ref{fig:headon} displays the result of a head-on collision of a pair of coaxial vortex rings that are initially moving in opposite directions and mutually annihilate as they collide head-on. This is not a new result and has been studied in detail in a different excitable medium by imposing axial symmetry directly within the numerical simulations \cite{BW}. Indeed, this process is equivalent to the annihilation of an isolated ring on encountering a no-flux boundary, where it effectively interacts with its mirror image. We present this head-on collision here so that we can compare with the equivalent threaded version in the next section.

The annihilation of one ring can be prevented if it is initialized with a radius that is larger than the other. This is illustrated by the example presented in Fig.~\ref{fig:head_on_swallow}, where again each ring is moving towards the other. The smaller lower blue ring moves inside the larger upper red ring but it is unable to escape. Effectively, the blue ring is swallowed by the red ring. More accurately, the small blue ring is destroyed by the waves emitted by the surrounding large red ring, as these waves squeeze the small blue ring below its stable radius. After the blue ring is destroyed, the red ring attains the stable ring radius. This process is consistent with the wavefront slapping mechanism discussed in \cite{Win_iop} in the context of stabilizing a knotted vortex string against contraction.  If the waves being produced by two vortex strings (or two different parts of the same string) have different frequencies then the higher frequency wavefronts will slap away the lower frequency source as the collision interface between wavefronts moves towards it. As we have seen in the previous section, the period of the minimal size ring is about $14\%$ above that of a large ring, hence the higher frequency wavefronts from the larger red ring slap the smaller blue ring.   

The swallowing phenomenon described above is ubiquitous in the interaction of unthreaded coaxial rings. Another situation in which swallowing can occur is when two rings are initialized close together with the same radius and moving in the same direction. An example is presented in Fig.~\ref{fig:swallow}, where both rings have a large initial radius and are moving up the $z$-axis. As in the interaction of vortex rings in fluids, the lower (blue) ring shrinks and therefore moves faster than the upper (red) ring, initiating an overtaking manoeuvre where the smaller (blue) ring attempts to pass through the larger (red) ring. However, the  overtaking manoeuvre is unsuccessful as the blue ring is swallowed by the surrounding red ring.

A successful overtaking manoeuvre is possible if both rings are initialized
sufficiently close together and the leading ring is created with a very large  radius that allows the following ring to pass through it.
An example is presented 
in Fig.~\ref{fig:overtake}, where both rings are moving up the $z$-axis and are
initialized close together with the upper (red) ring being created much larger than the slightly lower (blue) ring.
In this case the smaller (blue) lower ring is able to pass through the larger (red) upper ring and escape before the devastating tsunami wave arrives from the outer ring. This example shows that it is possible for overtaking to take place for unthreaded coaxial rings but the initial positions and sizes of the rings need to be finely tuned. The more generic outcome for unthreaded rings that encounter each other is that one ring is swallowed by the other.

Having highlighted the main processes that occur in the collision of coaxial rings, and demonstrated their fragility under multi-ring interactions, we now turn to the situation in which the coaxial rings are threaded on a vortex string.

\newpage
\section{Vortex rings on a string}\quad\label{sec:threaded}
In this section we investigate how the dynamics of coaxial vortex rings is modified by the introduction of a single vortex string that threads all the rings. The introduction of the threading string means that each vortex ring must be twisted, since each ring now has unit total twist, this being equal to the number of strings that thread it by the theorem \cite{WWS} mentioned earlier. Note that the threaded string breaks the axial symmetry of the fields in the unthreaded case, even for rings that are initially coaxial.

The phase profile for a vortex line that lies along the $z$-axis is simply
\be
\varphi_{line}=\arg\left(x+iy\right),
\ee
and by adding this contribution to the total phase profile we can introduce a threading string to the initial condition for any arrangement of coaxial rings.

\begin{figure}[ht]\begin{center}           
    \includegraphics[width=0.91\columnwidth]{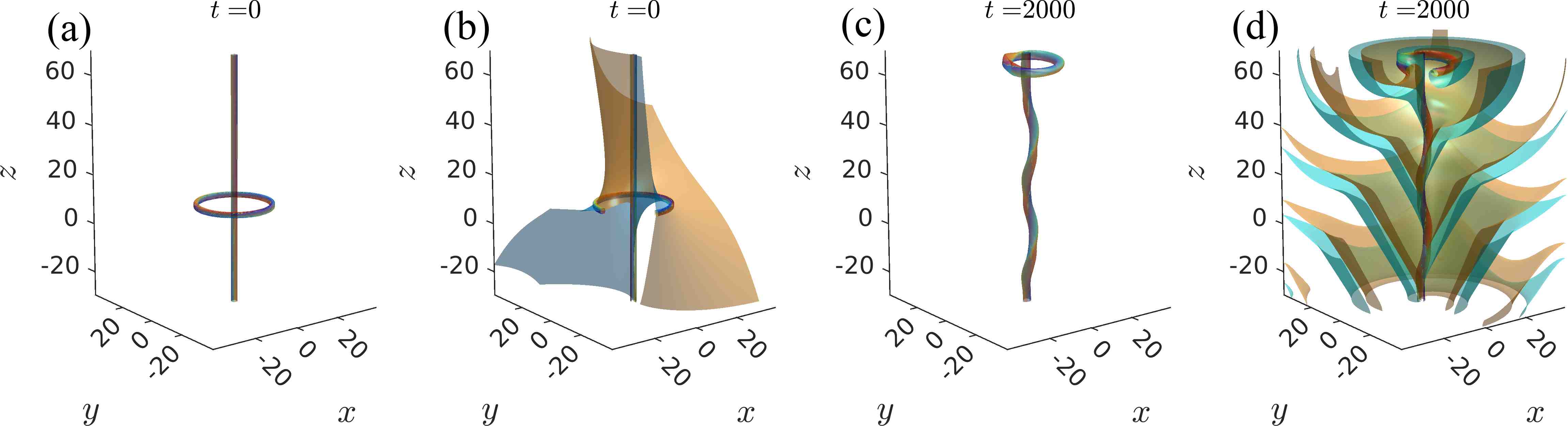}
    \caption{The initial condition for a single threaded vortex ring is shown in (a) and supplemented in (b) by the addition of the $u=0$ isosurface coloured by the value of the phase $\varphi$, with a quarter of the plot removed to aid visualization. In contrast to the unthreaded case, the ring is not destroyed at the no-flux boundary but remains there in a stable configuration (c), producing twisted wavefronts that form a kind of Archimedean screw (d).
  }
\label{fig:threaded_ring}
\end{center}\end{figure}
A single threaded ring has been investigated previously, see the review \cite{Win_siam} and references therein, with the result that there is a twist of the threaded string in the wake of the ring.
As specific details depend upon the parameter values and the system being studied, we reproduce the threaded ring example in Fig.~\ref{fig:threaded_ring},
for the parameter values used in the present paper, taking advantage of 
advances in computing capabilities to perform a more refined computation of this situation than was possible in early investigations \cite{Win_siam}.
The initial condition shown in Fig.~\ref{fig:threaded_ring}a already reveals the twist of the vortex ring, via the variation of the colouring along the ring, but to make this twist clearer we also present in Fig.~\ref{fig:threaded_ring}b the $u=0$ isosurface coloured by the value of the phase $\varphi$, with the front quarter of the plot removed to provide a better view.
The threaded ring stabilizes at a radius of roughly $10$, which is more than twice the unthreaded radius $R_\star$, and the ring drifts along the $z$-axis axis until it arrives at the no-flux boundary. However, in contrast to the unthreaded case, the ring is not destroyed at the no-flux boundary but remains there in a stable configuration, as shown at a late time in Fig.~\ref{fig:threaded_ring}c and Fig.~\ref{fig:threaded_ring}d, where the complicated twisting pattern of the phase of both the ring and the threaded string are visible. Note the twisting of the threaded string clearly visible in Fig.~\ref{fig:threaded_ring}c from the colour variation along the string, in addition to the helical conformation of the string.  The excitation wavefront is sandwiched between the two different coloured parts of the isosurface presented in Fig.~\ref{fig:threaded_ring}d.

\begin{figure}[ht]\begin{center}           
    \includegraphics[width=0.91\columnwidth]{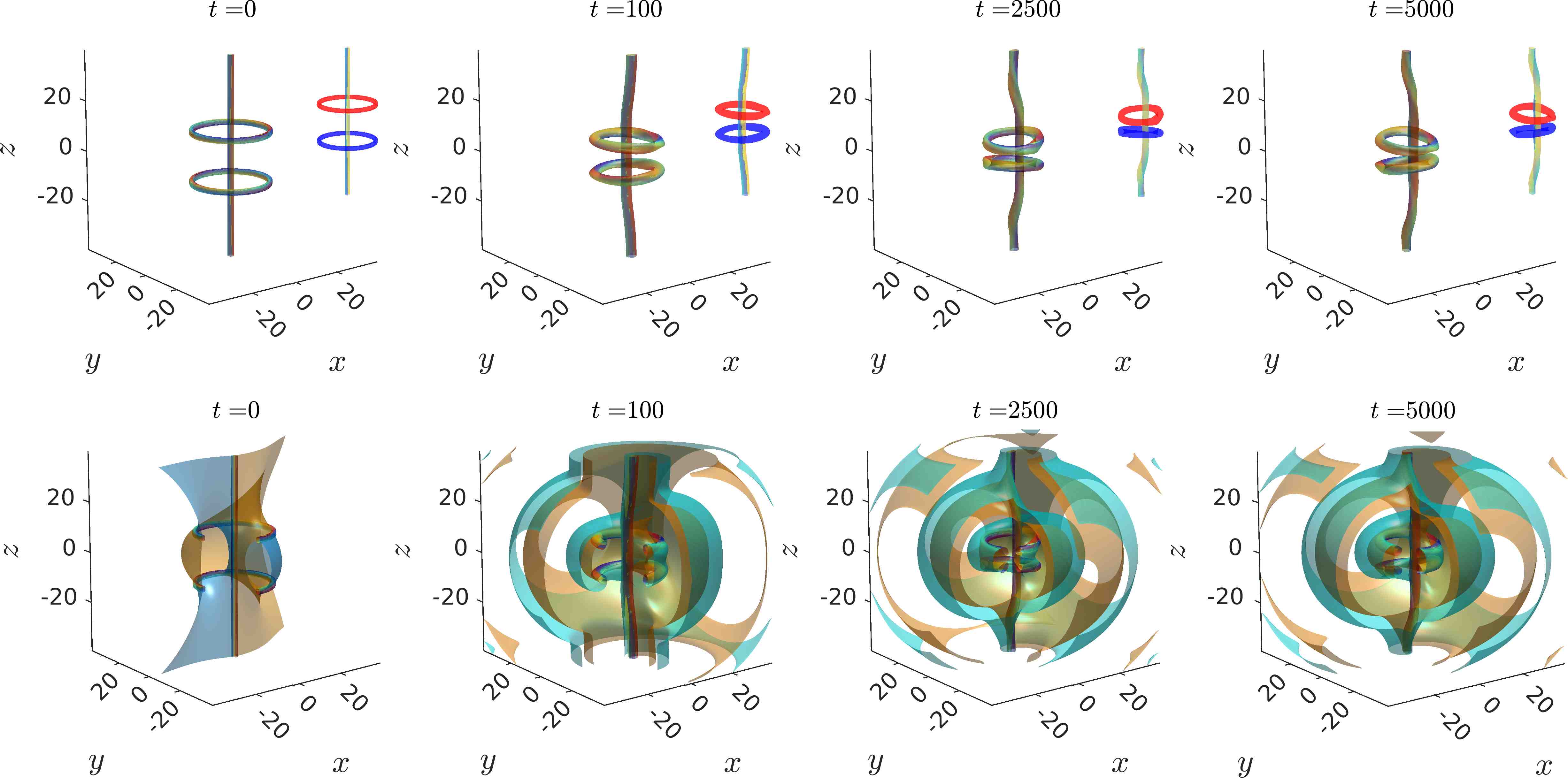}
    \caption{The head-on collision of a pair of symmetric threaded rings travelling in opposite directions. The threaded rings fail to annihilate but instead remain locked together.
The plots in the top row show the vortex ring and string cores, together with insets that colour the rings to aid identification. The plots in the bottom row reproduce those in the top row but without the insets and include $u=0$ isosurfaces coloured by the value of the phase, with a quarter of the plot removed to aid visualization. 
  }
\label{fig:blocking}
\end{center}\end{figure}

\begin{figure}[ht]\begin{center}           
    \includegraphics[width=0.91\columnwidth]{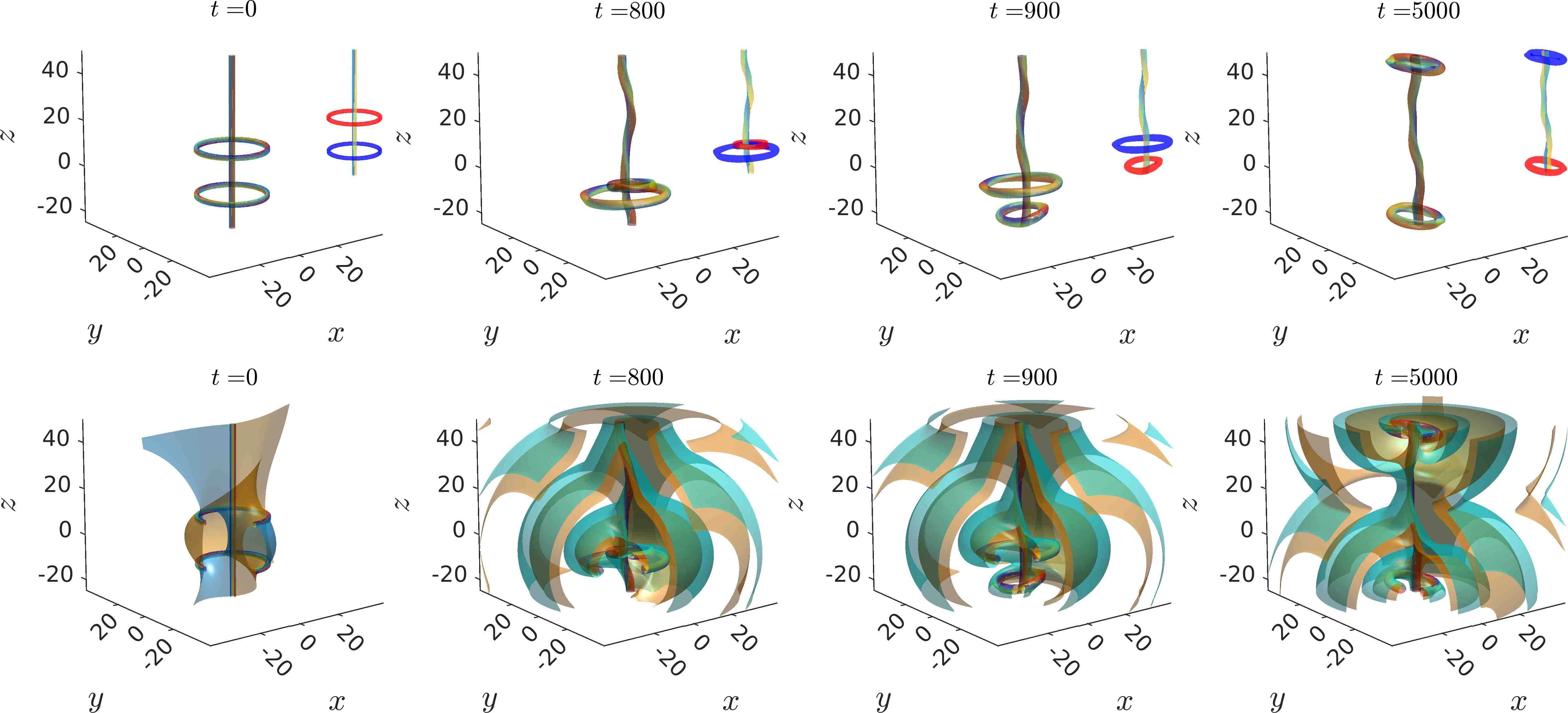}
    \caption{The head-on collision of a pair of asymmetric threaded rings travelling in opposite directions. The asymmetry is introduced by initializing the pair closer to the lower boundary of the medium than to the upper boundary. The upper ring (red in inset) shrinks and the lower ring (blue in inset) expands to allow one ring to pass through the other. After the interaction both rings separate and attain the stable size of a threaded ring and eventually settle at the upper and lower boundaries of the medium.
      The plots in the top row show the vortex ring and string cores, while the plots in the bottom row also include $u=0$ isosurfaces, coloured by the value of phase. 
  }
\label{fig:unblocking}
\end{center}\end{figure}
It has been observed both numerically and experimentally \cite{NW,ATE,TES} that in a range of excitable media the boundary of the medium provides an extra force that modifies the evolution of the radius of an approaching vortex ring. As we have seen in the previous section, for the system parameters studied in this paper, this extra force acts to reduce the radius of the ring and is sufficient to destroy an unthreaded ring as it approaches the boundary. However, the extra stability provided by a threaded string is enough to overcome this additional shrinking force and the threaded ring is able to sit snuggly at the no-flux boundary. 
As discussed earlier, a single ring at a no-flux boundary should be equivalent to a pair of rings, where one is the mirror image of the other. The situation presented in Fig.~\ref{fig:threaded_ring} can therefore be interpreted as the head-on collision of a pair of rings travelling in opposite directions. This is confirmed by the simulation presented in Fig.~\ref{fig:blocking}, where the head-on collision of a pair of symmetric rings is presented. In contrast to the unthreaded situation displayed in Fig.~\ref{fig:headon}, the threaded rings no longer annihilate but instead remain locked in a stalemate of mutual blocking.
Note the obvious lack of axial symmetry in this evolution due to the introduction of the threaded string.

The mutual blocking stalemate of identical rings can be resolved by introducing an asymmetry between the two rings. A novel way to introduce an asymmetry is to initialize the pair of rings closer to the lower boundary of the medium than to the upper boundary. As we have seen, a ring that is approaching a no-flux boundary experiences an additional shrinking force, therefore one expects that a ring moving away from a no-flux boundary will experience an additional expanding force, with the net result that the lower ring will initially become larger than the upper ring. This is indeed the case, as shown in Fig.~\ref{fig:unblocking}, where the two rings begin with the same size but the lower (blue) ring expands and the upper (red) ring shrinks and passes through the larger ring. 
Swallowing is thwarted by threading, so both rings survive the interaction
and separate to attain the stable size of a threaded ring. Both rings move in the direction of their original motion and eventually settle at the upper and lower boundaries of the medium. This example clearly demonstrates that threaded rings are more robust objects than their unthreaded counterparts, which is one of the main conclusions from this study.
The mechanism that prevents swallowing is subtle and appears to be a consequence of the fact that short range repulsion between all vortex cores prevents the inner threaded ring from entering a low frequency regime.
Recall from section \ref{sec:one} that an unthreaded ring with radius $R_\star$ has a period $1.14\,T$, compared with the period $T$ of larger rings, and is therefore susceptible to the slapping mechanism mentioned earlier. In contrast, for a threaded ring of minimal radius the period is computed to be $0.96\,T$ and is therefore slightly lower than the period $T$ of a large threaded ring. This reduction in period protects an inner threaded ring from slapping by a larger outer ring.  

\begin{figure}[ht]\begin{center}           
    \includegraphics[width=0.91\columnwidth]{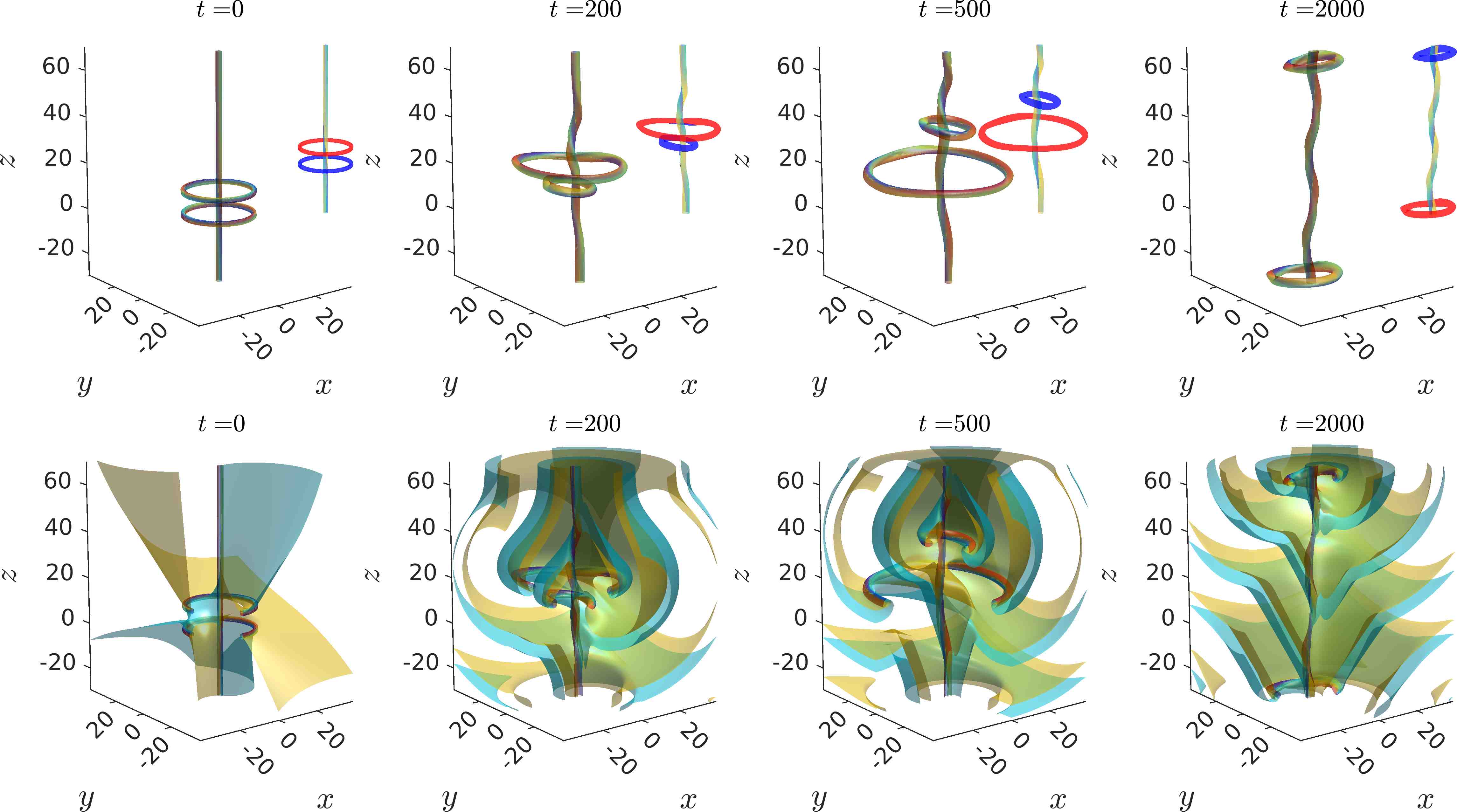}
    \caption{A threaded double ring interaction with insets that colour each ring to aid identification. The plots in the top row show the vortex ring and string cores, while the plots in the bottom row also include $u=0$ isosurfaces, coloured by the value of the phase.
      Two identical rings initially move up the $z$-axis. The lower (blue) ring shrinks and the upper (red) ring expands, so that the blue ring moves faster than the red ring and overtakes it by passing inside it. After the overtake, the blue ring attains the stable threaded ring size and continues to move up the $z$-axis until it reaches the upper boundary of the medium, where it remains. However, the red ring changes direction and moves down the $z$-axis, being pushed all the way down to the lower boundary of the medium by the twisted wavefronts. The red ring continues to shrink until it too attains the stable threaded ring size and it remains at the lower boundary.
  }
\label{fig:double}
\end{center}\end{figure}
The extra stability of rings provided by threading allows more complicated ring dynamics than in the unthreaded case, and indeed overtaking is now the generic outcome of ring interactions, rather than swallowing, as demonstrated in Fig.~\ref{fig:double}.
Both rings in Fig.~\ref{fig:double} are initially of equal size and are moving up the $z$-axis. The initial part of the evolution is similar to the unthreaded case and is familiar from the dynamics of vortex rings in fluids. The lower (blue) ring initially shrinks and the upper (red) ring initially expands. As smaller rings move faster, the blue ring overtakes the red ring by passing inside it. The blue ring is not swallowed and continues moving up the $z$-axis, attaining the stable threaded ring size and eventually reaching the upper boundary of the medium, where it remains. The aspect of the evolution that is unexpected, and differs from that of fluid vortex rings, is the motion of the red ring after the overtake. The direction of motion of the red ring is reversed by the overtaking manoeuvre and it now moves down the $z$-axis, despite its orientation remaining that of a ring that would naturally move up the $z$-axis. The red ring shrinks towards the stable threaded ring size as it moves down the $z$-axis and eventually it reaches the lower boundary of the medium, where it settles.

\begin{figure}[ht]\begin{center}           
    \includegraphics[width=0.7\columnwidth]{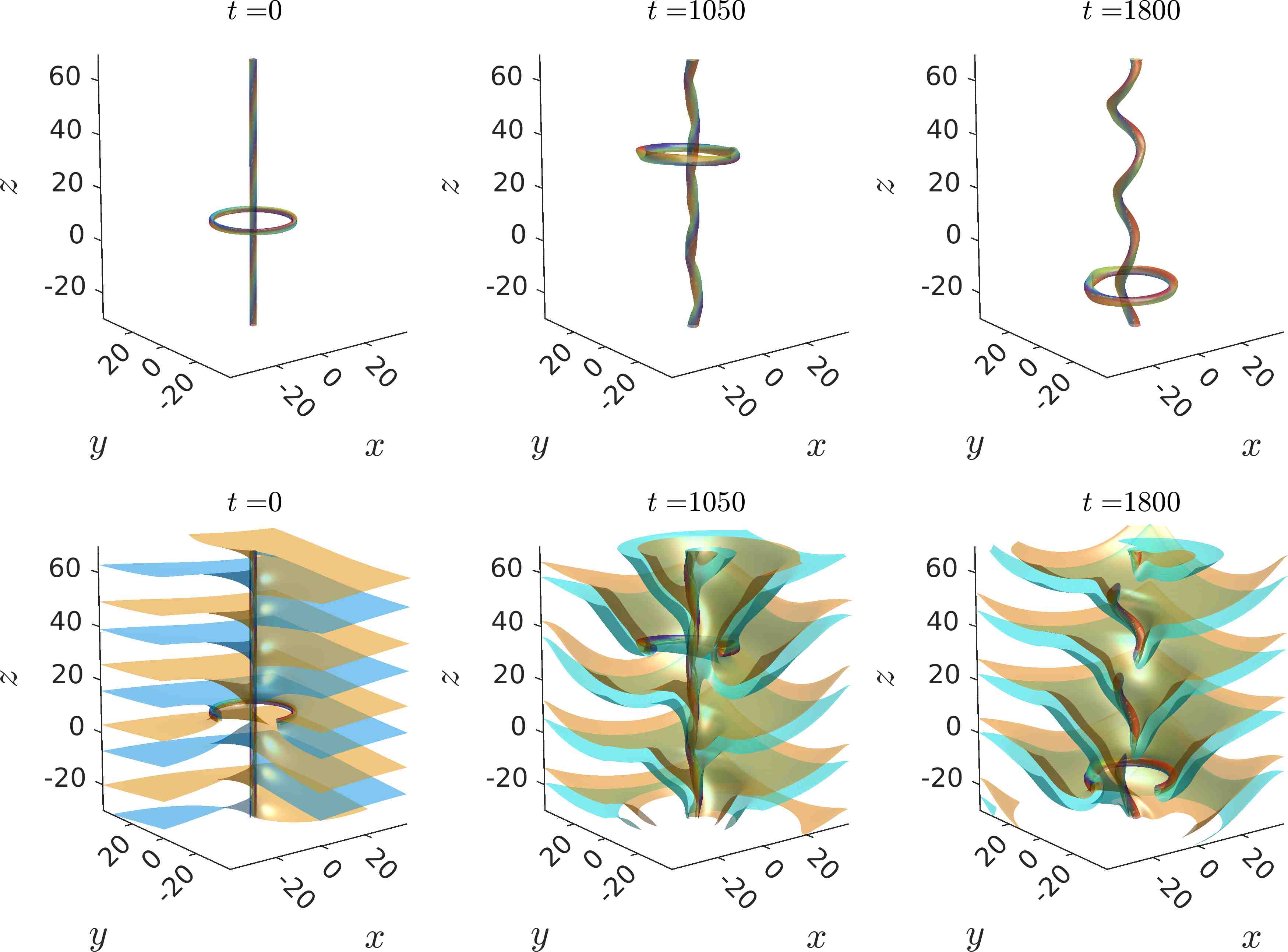}
    \caption{A vortex ring on a twisted threading string. The ring initially moves up the $z$-axis, as it would in the absence of twist, but the twisted waves from the string provide a kind of Archimedean screw that forces the ring to reverse its direction of motion and move down the $z$-axis.
 The plots in the top row show the vortex ring and string cores, while the plots in the bottom row also include $u=0$ isosurfaces coloured by the value of the phase,  with a quarter of the plot removed to aid visualization. 
  }
\label{fig:twisted}
\end{center}\end{figure}
The reversal of the direction of motion of a ring seen in this double ring overtaking evolution is surprising. A close examination of the phase of the threading string and the $u=0$ isosurfaces reveals that the mechanism producing a repulsive force in the wake of a ring is an indirect interaction via induced twisting of the threading vortex string. Note that in principle this twist can propagate an arbitrary distance along the string, so there is no limit on the range of this interaction.

To verify this mechanism we can induce a twisting of the threading string by hand, rather than due to the wake of a ring. To impose twist on a threading string we introduce a linear dependence on $z$ into the phase profile of the string and impose periodic boundary conditions in the $z$ direction, to prevent the string from untwisting. We then thread a single vortex ring onto this twisted string, as displayed in Fig.~\ref{fig:twisted}.
The orientation of the ring in Fig.~\ref{fig:twisted} is such that it would move up the $z$-axis in the absence of any twist in the threading string. Initially the ring does moves up the $z$-axis until the pattern of twisted waves emanating from the twisted threading string is established. As can be seen from the $u=0$ isosurfaces in Fig.~\ref{fig:twisted}, these twisted waves form a kind of  Archimedean screw that forces the ring to move down the $z$-axis and hence its direction of motion is reversed. As periodic boundary conditions are imposed in this simulation the ring moves through the bottom boundary and reappears at the top boundary, continually moving down the $z$-axis. Note the similarity in the wave pattern produced by the twisted string in Fig.~\ref{fig:twisted} and the wake of a threaded ring in Fig.~\ref{fig:threaded_ring}. In this example the twist rate of the string is sufficient to reverse the direction of motion of the ring but, as expected, a lower twist rate has a reduced effect. Indeed, we have found a critical twist rate so that the natural motion of the vortex ring up the $z$-axis is exactly balanced by the strength of the Archimedean screw and the ring remains at a constant position.

\begin{figure}[ht]\begin{center}           
    \includegraphics[width=0.91\columnwidth]{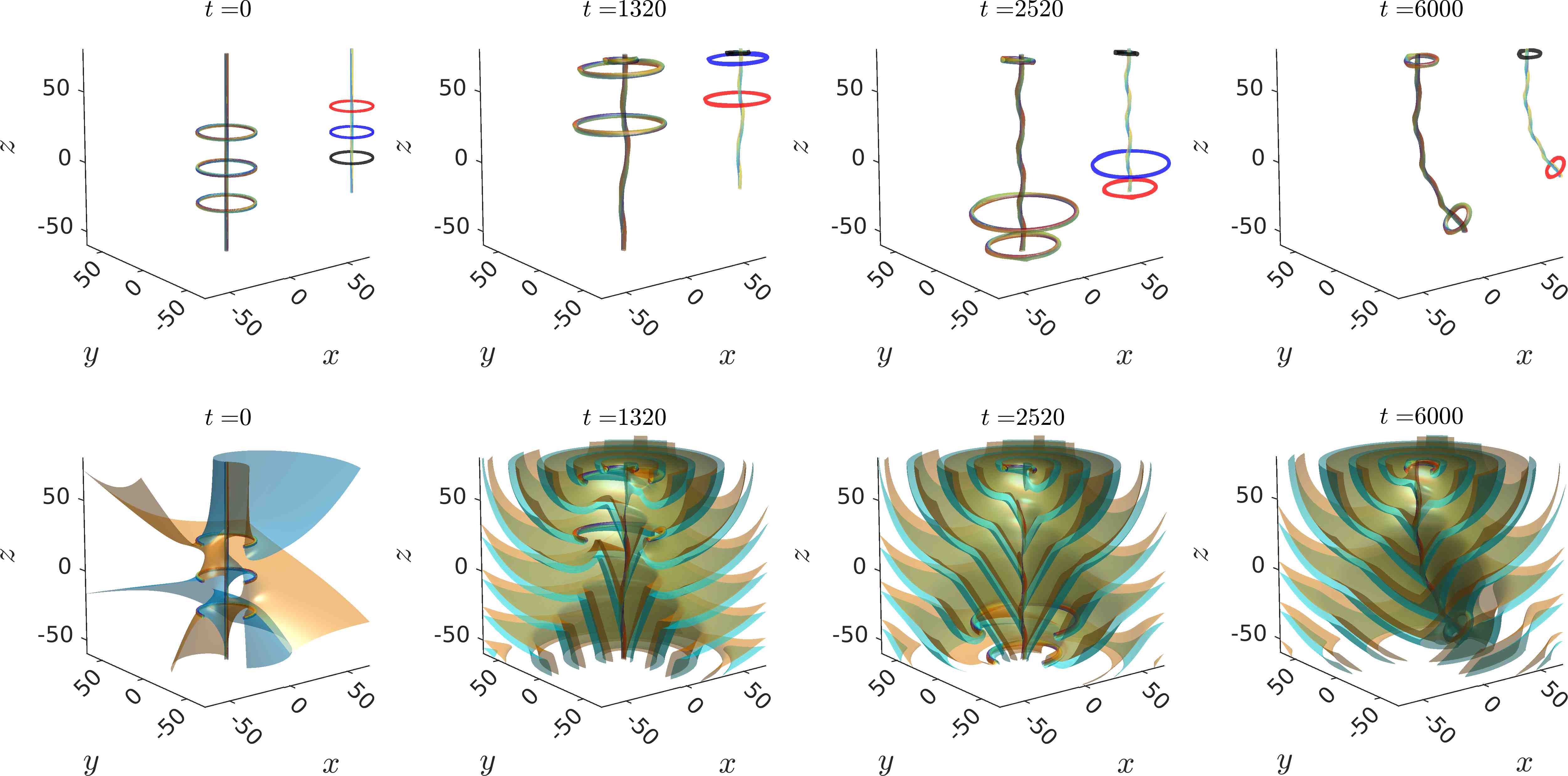}
    \caption{A threaded triple ring interaction with insets that colour each ring to aid identification. The plots in the top row show the vortex ring and string cores, while the plots in the bottom row also include $u=0$ isosurfaces coloured by the value of the phase.
Three identical rings initially move up the $z$-axis but a series of overtakes results in a complete reversal of the original order. The top ring then sits at the upper boundary of the medium and the other two rings are pushed down to the lower boundary of the medium where a final overtake occurs, resulting in the annihilation of the lowest ring as it is pushed into the boundary.     
  }
\label{fig:triple}
\end{center}\end{figure}
We have seen that the interaction of a pair of threaded rings yields interesting dynamics, and even more exotic behaviour is found if more rings are included.
The evolution of several threaded rings is demonstrated in the triple ring simulation presented in Fig.~\ref{fig:triple}.
     All three rings in Fig.~\ref{fig:triple} are initially of equal size and are moving up the $z$-axis. First, the upper (red) ring expands to allow both the middle (blue) and lower (black) rings to overtake it. Next the blue ring expands to allow the black ring to overtake it, resulting in a complete reversal of the original order of the triple rings. The black ring then sits at the upper boundary of the medium and induces a twisting of the threaded vortex string in its wake. The twisted waves in the wake of the black ring are sufficient to reverse the direction of motion of the blue and red rings, which are now pushed down the $z$-axis, despite their orientations remaining those of rings that would naturally move up the $z$-axis. The red and blue rings move as a pair down the $z$-axis until the red ring reaches the lower boundary of the medium where it settles. The blue ring expands as it approaches the red ring and surrounds it, as it too sits at the boundary. After a short period of this concentric configuration the inner red ring moves slightly away from the boundary so that the larger blue ring is now the lowest ring. The wake of the red ring then pushes the blue ring into the lower boundary where it annihilates with its mirror image. Thus threading is sufficient to prevent annihilation of an isolated ring moving towards the medium boundary but even the extra stability provided by threading can be overwhelmed if sufficient forces are present due to interactions with waves generated by other rings.

     As we have demonstrated with several examples, threaded rings are more stable than their unthreaded counterparts and behave more like fluid vortex rings in terms of their ability to overtake. Vortex rings in fluids can perform the famous leapfrogging motion discussed by Helmholtz \cite{Helm} in the mid-nineteenth century, where the lower ring shrinks and overtakes the upper ring but then expands so that the original configuration is recovered but with an exchange of the two rings, allowing the overtaking process to repeat. The same leapfrogging locomotion is found in other systems with vortex rings, such as in Landau-Lifshitz simulations of the magnetization in a ferromagnetic medium \cite{NS}, but we have been unable to find any evidence of leapfrogging in our studies of either threaded or unthreaded vortex rings in excitable media. This is presumably a result of the extra interaction between rings via the induced twisting of the threaded string, which forces the change of direction of an overtaken larger ring and means that the configuration after overtaking is never equivalent to the configuration before overtaking (even up to an exchange of rings).
     In any case it is clear that although threaded rings share some features with vortex rings in fluids there are certainly important differences.

\section{Conclusion}\quad
Motivated by recent results on knotted vortex strings and untangling in excitable media, we have investigated the influence of linking on vortex string interactions by considering the situation of coaxial vortex rings, with minimal linking introduced via a single threading vortex string. This setup allows for a controlled study of linking modified interactions because the evolution can be compared with the same scenario of coaxial rings in the absence of the threading string.

We have shown that linking provides an increased stability, with the fragility of unthreaded rings being replaced by a robustness that allows threaded rings to survive overtaking processes. We attribute this to the repulsion between all vortex cores, that prevents threaded rings from entering the low frequency regime where they are vulnerable to slapping.  
This improved stability yields a much richer range of dynamical events and reveals that the behaviour of excitable threaded vortex rings is much closer to fluid vortex rings than their unthreaded counterparts, in that swallowing is not a generic outcome. The signature overtaking process of fluid vortex rings is reproduced by threaded excitable rings, in contrast to the annihilation found in the unthreaded case. However, there are also new features due to threading that are unexpected, and include the reversal of the direction of motion of a ring via induced twisting of the threading vortex string in the wake of another ring. If any type of effective string model is to reproduce the complicated dynamics found in knotted vortex strings then it must first be able to match the examples of evolution presented in the present paper, which represent the simplest stage to inspect the influence of linking on interactions. As we have seen, even in the simple situation of coaxial rings with minimal linking, the evolution is complex and will certainly provide a significant challenge to effective string models to capture such complicated behaviour.

There has been impressive and detailed recent experimental work \cite{ATE,TES} on vortex rings in the excitable Belousov-Zhabotinsky medium. This experimental work has been able to study the influence of the medium boundary on vortex ring dynamics and it would be very interesting if similar work could be undertaken to investigate coaxial rings and to demonstrate some of the qualitative new features from threading that are predicted in the present paper. Although the numerical simulations presented are within the FitzHugh-Nagumo medium we have also performed computations using the three-component Oregonator equations that model the Belousov-Zhabotinsky medium and found similar results. We therefore expect that our findings are amenable to experimental validation.  

\section*{Acknowledgements}
\noindent  
This work is funded by the
Leverhulme Trust Research Programme Grant RP2013-K-009, SPOCK: Scientific Properties Of Complex Knots.

\end{document}